\documentclass[lettersize,journal]{IEEEtran}
\usepackage{amsmath,amsfonts}
\usepackage{amssymb}
\usepackage{booktabs}
\usepackage{multirow}
\usepackage{pifont} 

\usepackage{algorithm}
\usepackage{algpseudocode}
\usepackage{array}
\usepackage[caption=false,font=normalsize,labelfont=sf,textfont=sf]{subfig}
\usepackage{textcomp}
\usepackage{stfloats}
\usepackage{url}
\usepackage{verbatim}
\usepackage{graphicx}
 
\usepackage{xcolor}
\usepackage{microtype}
\usepackage{wrapfig}
\usepackage{lipsum} 
\usepackage{float}

\begin{document}

\title{$C^2$AV-TSE: Context and Confidence-aware Audio Visual Target Speaker Extraction}

\author{Wenxuan Wu, Xueyuan Chen, Shuai Wang, Member, IEEE, Jiadong Wang, Lingwei Meng,\\ Xixin Wu, Member, IEEE,
Helen Meng, Fellow, IEEE, Haizhou Li, Fellow, IEEE





}



\maketitle

\begin{abstract}

Audio-Visual Target Speaker Extraction (AV-TSE) aims to mimic the human ability to enhance auditory perception using visual cues. Although numerous models have been proposed recently, most of them estimate target signals by primarily relying on local dependencies within acoustic features, underutilizing the human-like capacity to infer unclear parts of speech through contextual information. This limitation results in not only suboptimal performance but also inconsistent extraction quality across the utterance, with some segments exhibiting poor quality or inadequate suppression of interfering speakers. To close this gap, we propose a model-agnostic strategy called the Mask-And-Recover (MAR). It integrates both inter- and intra-modality contextual correlations to enable global inference within extraction modules. Additionally, to better target challenging parts within each sample, we introduce a Fine-grained Confidence Score (FCS) model to assess extraction quality and guide extraction modules to emphasize improvement on low-quality segments. To validate the effectiveness of our proposed model-agnostic training paradigm, six popular AV-TSE backbones were adopted for evaluation on the VoxCeleb2 dataset, demonstrating consistent performance improvements across various metrics.

\end{abstract}


 \vspace{-10pt}
\section{Introduction}
Audio-visual target speaker extraction (AV-TSE) aims to isolate the target speaker's voice from a mixed audio source conditioned on visual cues. This process is inspired by the human selective listening mechanism, which allows the individuals to concentrate on specific speakers in a noisy environment, such as the classic ``cocktail party'' scenario~\cite{brain,select-listen}. 
Currently, most AV-TSE models employ lip movement as visual cues because they are informative, robust to acoustic noises and highly correlated with speech content \cite{muse,av-sepformer,usev,tdse,wu2024target_cvpr,10650095,ImagineNET}.
Such systems typically focus on obtaining efficient audio-visual synchronization cues and corresponding effective integration methods \cite{muse,av-sepformer,usev,tdse,wu2024target_cvpr,10650095,ImagineNET}, facilitating the alignment of audio-visual information.


Most current AV-TSE systems primarily rely on the synchronization between audio and visual modalities for extraction. However, beyond such synchronization, the human brain employs additional mechanisms to track a target speaker’s voice, such as predicting upcoming words or recalling prior memories of the speaker. By capturing contextual correlation from the target speaker's utterances, the human brain may focus on the target speaker's voice more effectively while ignoring interfering speakers' utterances.


In previous studies on speech separation, researchers have attempted to leverage contextual information by integrating Automatic Speech Recognition (ASR) modules with separator \cite{Wu2019ImprovingSE, e2easr_separation, li20q_interspeech}. 
However, this approach predominantly captures the content-related information, such as semantics and grammar, while neglecting environmental acoustic context and paralinguistic cues, such as the target speaker's emotion and speech rate,
 which are not visible in ASR transcripts.
These forms of contextual information provide valuable cues for target speaker extraction, highlighting the importance of developing robust contextual modeling strategies that can effectively capture and utilize these diverse contexts.
Recently, the Masked Language Modeling (MLM) strategy has gained popularity due to its strong contextual modeling ability \cite{BERT}. Similarly, Masked Acoustic Modeling (MAM) has been effectively applied to many speech foundation models, such as wav2vec 2.0 \cite{wav2vec2}, HuBERT \cite{HuBERT}, and SSAST \cite{Ssast}. Leveraging this self-supervised pre-training strategy, these models produce speech representations enriched with contextual information and have consistently demonstrated superior performance across a wide range of speech-related downstream tasks \cite{wav2vec2, HuBERT}.

\begin{figure}[t]
\centering
 \includegraphics[scale=0.7]{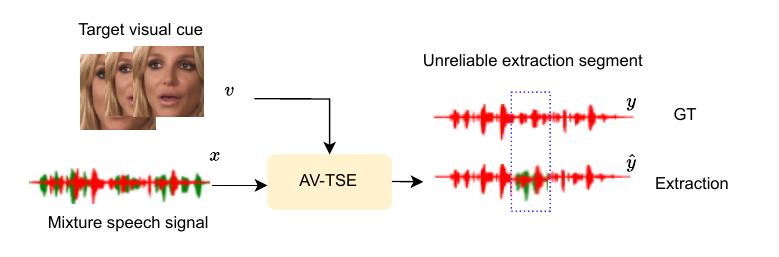}
\caption{The unreliable segments are from AV-TSE results. $v$ and $x$ denote the target visual cue and mixture speech signal, respectively. One unreliable extraction segment is indicated with dotted rectangles, where the interfering speech signal is in light green, and the target speech signal is in red.}
\label{Figure1}
\end{figure}

In this paper, instead of explicitly integrating ASR modules, we propose a two-stage contextual information-enhanced fine-tuning approach, featuring a model-agnostic Mask-And-Recover (MAR) training strategy. This strategy leverages both content and acoustic context from audio-visual modalities to recover masked frames, enabling the extractor to learn inter-modality context from mixtures and intra-modality context from visual cues.
The first phase of fine-tuning, referred to as ``\emph{global fine-tuning}'', aims to improve the overall extraction performance. In this phase, the mixture segments are randomly masked with the proposed MAR strategy. In the second phase, we focus specifically on mixture segments with poor extraction performance. This strategy is motivated by the observation that, within a given speech utterance, the performance across different segments is often inconsistent. Such inconsistencies can arise due to model uncertainty or inherent characteristics of the input mixture data.
To identify the segments requiring special attention, termed “unreliable extraction segments” in this study (see Fig. \ref{Figure1}), we propose a confidence-based measurement called the Fine-grained Confidence Score (FCS) prediction model. Unlike generating utterance-level confidence scores, the FCS model evaluates the reliability of individual frames in the extracted speech. Once unreliable segments are detected, they can either be refined with the supervised signal or masked and subsequently improved using the proposed MAR strategy. By incorporating such confidence information into the AV-TSE system, the second phase is termed as ``\emph{confidence-aware fine-tuning}''.

The contributions of this paper can be summarised as follows:
\begin{itemize}
 \item We highlight the importance of intra- and inter-modality contextual information for AV-TSE systems and propose the MAR strategy, which effectively integrates visual context with target speech context during the extraction process.

 \item We introduce the FCS model, which enables the automatic detection of unreliable extraction segments. To the best of our knowledge, this is the first implementation explicitly designed to locate and address unreliable TSE outputs automatically. 

 \item We propose the C$^2$AV-TSE framework, a two-stage fine-tuning approach that employs a progressive optimization strategy transitioning from global to local and from general to hard. This framework can be applied to any existing AV-TSE system.

 \item {Comprehensive experiments conducted across six AV-TSE models demonstrate consistent improvements across various evaluation metrics.}
\end{itemize}
 

\section{Related work}

\subsection{Contextual Information Modeling}

Speech context encompasses rich information, including speaker characteristics, linguistic structure, and semantic features. This contextual information is crucial for enhancing performance across various speech-related tasks. In current literature, speech contextual features are often utilized in conjunction with masking strategies.

One notable application of such strategies is in speech editing tasks, where regions requiring edits are masked as inputs. For instance, CampNet \cite{CAMPNET} employs a masking strategy for end-to-end text-based speech editing. Similarly, MaskedSpeech \cite{zhang23n_interspeech} uses a masking strategy in speech synthesis tasks to capture cross-utterance semantic features at different resolutions. To achieve high-quality speech representations, $A^3T$ \cite{A3T} integrates text inputs and acoustic-text alignment features to reconstruct masked acoustic signals during pre-training.

In the ASR task, speech context also plays a significant role. In \cite{Cui2023TowardsEA}, a compact contextual representation from cross utterances is extracted to improve streaming ASR performance. Furthermore, \cite{Robertson2023BiggerIN} investigates the impact of contextual size on speech pre-training using contrastive predictive coding (CPC) \cite{CPC}, revealing that larger contextual sizes do not necessarily benefit ASR. Additionally, MAM \cite{Chen2020MAMMA} and FAT-MLM \cite{Zheng2021FusedAA} implement masking strategies to effectively learn speech context for speech translation tasks.

In speech separation tasks, \cite{li20q_interspeech} proposes adding a contextual embedding prediction model to learn contextual information from the mixture, utilizing audio, visual, and extracted context as separation conditions. However, this approach requires training the contextual embedding model on ASR tasks, which presents limitations in practical applications. To address this challenge, we propose a mask-and-recovery strategy that enables Audio-Visual Target Speaker Extraction (AV-TSE) models to learn speech context without needing an embedded contextual prediction model or ASR model integration.

\subsection{Hard Sample Mining}

Generally, a ``hard sample'' in machine learning refers to a data sample near the model decision boundary that may lead to an incorrect prediction \cite {Robinson2020ContrastiveLW}. Hard sample mining is a challenging yet critical task in many research fields \cite {Lin2017FocalLF}, leading to better generalization and refined decision boundaries.

In previous studies, hard sample mining algorithms have succeeded in many fields. In computer vision such as the face recognition task, the authors found explicitly mining the hard triplets can lead to faster model convergence \cite{Schroff_2015_CVPR}. Similarly, in \cite{Lin2017FocalLF}, the author suggested that hard samples are crucial to limit one-stage dense objective detector performance. To solve this, the focal loss is further proposed to improve detection accuracy. In speech processing, in \cite{Hou2020MiningEN}, to address the class imbalance challenge in the keyword Spotting task, the author proposed an algorithm for mining hard samples and controlling the ratio between positive and negative samples. 

More recently, hard sample mining algorithms have also been explored in conjunction with self-supervised and generative models to enhance model performance. In \cite{Hard_Patches_Mining}, the authors explored hard sample mining algorithms with Masked Visual Modeling (MVM) to improve pre-training effectiveness. Similarly, \cite{sahin2024enhancing} proposed a generative approach to produce hard negative samples, thereby enhancing the distance between positive and negative samples.

Inspired by previous studies, we explore the strategy to overcome such hard samples to improve extraction performance. 
Specifically, in this paper, we define the hard samples as \emph{unreliable extracted speech segments}. The proposed FCS model, serving as a hard sample estimator, aims to identify these challenging samples. By leveraging the FCS estimator, we can focus more on these error-prone samples and conduct targeted training, thereby enhancing overall performance.




\section{Contextual Information Modeling}

As mentioned in the Introduction, contextual information such as speech character, linguistic structure, and semantics have been proven effective in many speech-related tasks. In this section, we will introduce the details of our proposed Mask-And-Recover (MAR) strategy and explain how contextual information is integrated into the separator. 


 \begin{figure}[!t]
\centering
\includegraphics[scale=0.8]{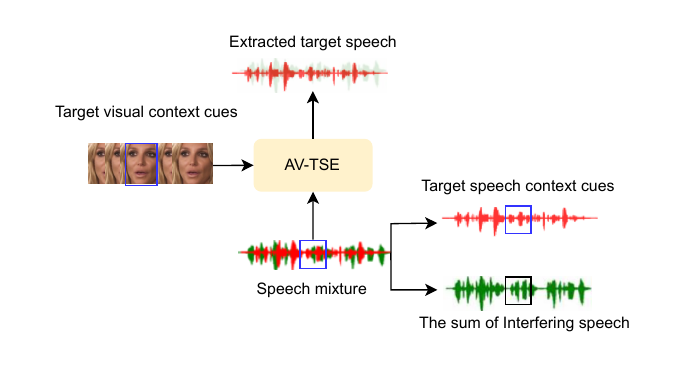}
\label{CONFIDENCE-TSE}
\vspace{-25pt}
\caption{Contextual cues in AV-TSE. The current frame is denoted by a blue rectangle. In addition to the corresponding visual cue in the blue rectangle, the target visual context and target speech context also serve as additional cues for the extraction. In contrast, the green signal denotes the sum of interfering speech, which may harm the extraction performance.}
\vspace{-15pt}
\label{context_cues}
\end{figure}

\subsection{Inter- and Intra-Modality Contextual Information}

In this section, we will introduce the intra-modality contextual information from a mixture of speech signals, and inter-modality context from target lip movement, respectively. Then we will introduce the proposed MAR strategy, which captures these contextual information and integrates them into the separator.

\subsubsection{Intra-modality Cues From Target Speech In Mixture}
Speech context provides rich semantic information, linguistic structure, paralinguistic features, and speaker characteristics, which could serve as useful cues for extraction. As shown in Fig. \ref{context_cues} suppose the speech mixture consists of the target speech signal and the sum of interfering speech signals. The former is represented in red, while the latter is denoted in green. To extract the target speech segment from the mixture, displayed in the blue box, the remaining parts could provide such contextual cues, indicated by the dotted line.
However, one challenge is that the interfering speech context, considered irrelevant to the target speech, may negatively affect the extraction performance. 
 
\subsubsection{Inter-modality Cues From Target Lip Movements}
As illustrated on the left side of Fig. \ref{context_cues}, the extraction of the target speech frame in the blue box relies significantly on the corresponding target lip movement within the same blue box, which offers crucial viseme information and serves as the most direct cue. In addition to this, the target lip movements throughout the same utterance provide valuable visual contextual cues, aiding in identifying the target speaker.

\subsubsection{Mask-And-Recover Strategy}
\begin{figure}[!t]
\centering
\includegraphics[scale=0.6]{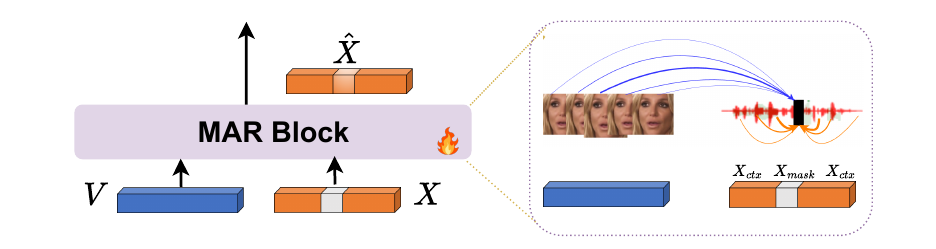}%
\label{CONFIDENCE-TSE}
\vspace{-25pt}
\caption{Illustration of MAR strategy. The input to the speaker extractor includes intact visual cues $v$ and masked mixture speech signal $x$. The output of the speaker extractor includes extracted speech embedding $X$ (shown in orange) and corresponding visual cue embedding $V$ (shown in blue). To recover the masked region $X_{mask}$, both the intra-modality context from target speech context $X_{ctx}$ 
as well as inter-modality context from $V$ will contribute. Here, the temporal synchronized visual cue of $X_{mask}$ serves as a direct visual cue, and the remaining visual frames serve as visual contextual cues. To distinguish different levels of contribution, the relevance of the context is represented by curves of varying thickness. 
By modeling both types of contextual information during extraction, the learned contextual correlation will be injected into the speaker extractor as additional extraction cues.}
\label{MAR strategy}
\end{figure}
 As shown in Fig. \ref{MAR strategy}, to fully leverage contextual cues from target speech and target lip movement, we propose the Mask-And-Recover (MAR) strategy. Specifically, we mask certain frames of the speech mixture as AV-TSE model input. 
The speech mixture is first processed through the speech encoder, while the target visual cue is fed into the visual encoder and visual adapter. The outputs of these components are then passed to the speaker extractor, which produces an estimated target speech embedding, denoted as $X$. Note that in some AV-TSE systems, the visual embedding $V$ is refined within the speaker extractor; therefore, we use the visual embedding $V$ from the output of the speaker extractor rather than from the output of the visual adapter. After this, we concatenate $X$ and $V$ and input them to the MAR block, as demonstrated in Fig. \ref{MAR strategy}. To reconstruct the masked region $X_{mask}$, the extracted target speech context provides intra-modality correlations, represented by the orange curve in Fig.~\ref{MAR strategy}. The significance of speech context varies with the temporal proximity to the masked region, as indicated by varying line thicknesses. Similarly, target lip movements offer inter-modality correlations, depicted by the blue curve. However, due to the coarser resolution of visual data compared to speech, the visual context contributes less significantly than directly synchronized visual cues. The output of MAR block $\hat{X}$ is expected to contain rich contextual information and well-extracted target speech.

It is important to note that by incorporating the MAR block and tailored loss functions, the contextual information will be integrated into the speaker extractor, serving as additional extraction cues. This method contrasts with the cascade approach, which involves pre-extracting the target speech and subsequently refining it separately with MAR. Instead, we optimize the MAR block with the speaker extractor in a joint manner. 

\section{Confidence Measurement}

To the best of our knowledge, we are the first to utilize confidence scores to identify unreliable extraction segments in TSE outputs. As no existing annotated datasets are directly applicable, we address this challenge through a multi-step approach. We first analyze the potential target speech extraction states and then simulate a dataset with paired confidence scores to approximate the main components of real TSE outputs. Then, we train a model to predict fine-grained confidence scores with the simulated dataset. 

\subsubsection{Potential States of TSE Outputs}

Commonly, the input mixture can be regarded as a combination of target speech $y$ and interfering speech $z$. We hypothesize that the AV-TSE system's output can also be approximated as a weighted combination of these same components, although with modified mixing ratios. Following this hypothesis, the simulated extraction output can be formulated as Equation~\ref{eq2}.

\text{For a given waveform sample at time $t$}, 
\begin{equation}
\begin{split}
\hat{y}_t = \alpha \cdot y_t + \beta \cdot z_t, \\
\alpha \in [0,1], \beta \in [0,1]
\end{split}
\label{eq2}
\end{equation}
where $\alpha$ and $\beta$ are scale factors representing the remaining degree of target speech $y_t$ and interfering speech $z_t$ from the mixture, respectively. According to Equation \ref{eq2}, $\hat{y}_t$ can be represented within a 2D coordinate system with $\alpha$ and $\beta$ as axes. Specifically, the coordinate point $(1, 0)$ represents perfect extraction at time $t$, where $\hat{y}_t = y_t$. Conversely, $(0, 1)$ indicates complete target confusion \cite{target-speaker-confusion}, where the system has entirely extracted the interfering speech instead of the target speech.
Moreover, $(1,1)$ represents the unprocessed input mixture, while $(0,0)$ represents complete silence, which rarely occurs in practice.

\subsubsection{TSE Output Simulation}

According to the aforementioned analysis, we approximate the TSE result by controlling the remaining proportions of target and interfering speech components. Suppose $\mathcal{Y}$ is all target speech utterances, and $\mathcal{Z}$ is all interfering speech utterances, to better approximate real extraction results, we consider several key parameters: the number of unreliable segments per utterance $N_{max}$, the duration of each unreliable segment $g$, and the sampling strategy. 
The detailed simulation procedure is presented in Algorithm \ref{algorithm1}.
\begin{algorithm}
\caption{TSE output simulation}
\label{algorithm1}
\begin{algorithmic}[1]
\State \textbf{Input:} $(\mathcal{Y}, \mathcal{Z}, \alpha, \beta, g, N_{max})$
 \State \textbf{Output:} $\tilde{\mathcal{Y}}$- Simulated target speech utterances
 
 \For{$y \in \mathcal{Y}$} 
 
 \State Randomly sample an interfering speech utterance $z \in \mathcal{Z}$ 
  \State Initialize simulated target speech utterance $\tilde{y} = y $
 \State Sample a value for $N$, $N \sim U(0,N_{max})$
  \For{$n \gets 1$ to $N$}
 
  \State Sample a random start time $t_n \sim U(0, T - g )$
  
  \State $\tilde{y}[t_n:t_n+g] = \alpha \cdot y[t_n:t_n+g] + \beta \cdot z[t_n:t_n+g]$
  \EndFor

 \State Append $\tilde{y}$ to $\tilde{\mathcal{Y}}$
 \EndFor
 \State \Return $\tilde{\mathcal{Y}}$ 
\end{algorithmic}
\end{algorithm} 

Note that when training the FCS model with BCE loss, the unreliable segments simulated with $\alpha$ 
 and $\beta$ are labeled with $1$, while the clean speech segments are labeled with $0$.


\subsubsection{Fine-Grained Confidence Score Prediction Model}
\begin{figure}[!htb]
\centering
 \centering
\includegraphics[scale=1]{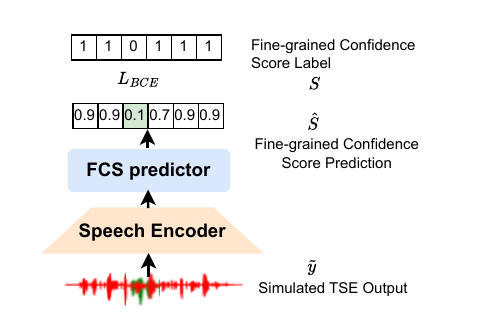}
\caption{Fine-Grained Confidence Score (FCS) Prediction model}

\label{FCSmodel}
\end{figure}

 The fine-grained confidence score prediction (FCS) model employs a straightforward architecture. As shown in Fig. \ref{FCSmodel}, the simulated TSE utterance $\tilde{y}_{[1:T]}$ will be first processed by a speech encoder and then be fed into the FCS predictor. The speech encoder shares a similar architecture with the speech encoder in the AV-TSE system but with a different kernel size and stride. The FCS predictor starts with a linear layer, followed by a transformer block, aiming to capture the temporal acoustic variance feature. Finally, another linear layer and a subsequent Sigmoid activation function are applied to predict confidence scores $\hat{S}_{[1:T']}$,where $T' < T$.
 
\vspace{-15pt}
\begin{equation}
L_{BCE}(\hat{S}_{[1:T']},S_{[1:T']}) = -\sum_{t=1}^{T'} S_t \cdot \log(\hat{S}_t)+(1-S_t) \cdot \log(1-\hat{S}_t)
\end{equation}
The FCS model is trained with the Binary Cross Entropy (BCE) loss.

\section{Context and confidence aware fine-tuning}
\begin{figure*}[!t]
 
\includegraphics[scale=0.3]
{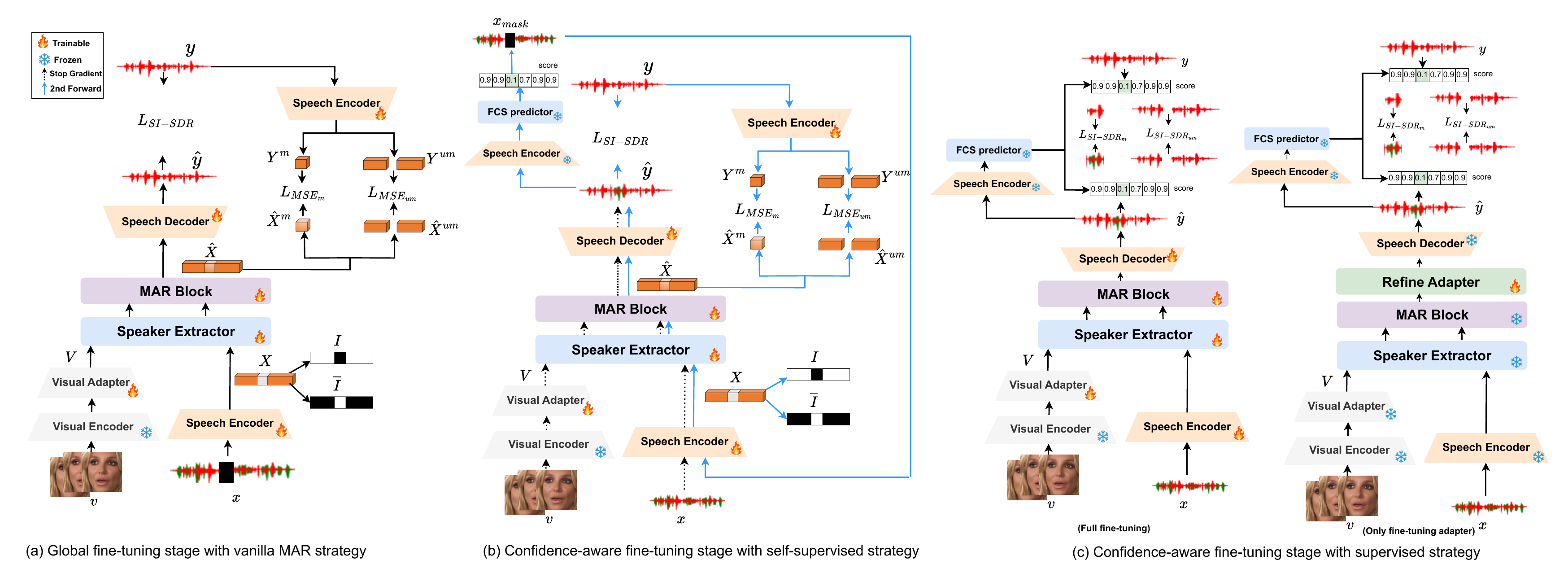}%
\caption{Illustration of the proposed two-stage fine-tuning strategy. \textbf{Stage 1}: (a) Global fine-tuning, utilizing the vanilla MAR strategy with randomly masked mixture speech $x$ and intact visual cue $v$. All modules except the MAR block are initiated from the pre-trained AV-TSE model. All modules will be fine-tuned excluding the visual encoder. $I, \overline{I}$ denote the mask automatically detected from $x$. \textbf{Stage 2}: (b, c) Confidence-aware fine-tuning. All modules are initiated from stage 1. A frozen pre-trained FCS model is integrated to detect unreliable extraction segments. (b) Self-supervised fine-tuning, similar to Stage 1, but mask segments of $x$ based on predicted confidence scores. The first forward, shown in the dotted line, stops the gradient to infer the masked region. The second forward, shown in blue, is self-supervised fine-tuning with masked mixture input. (c) Supervised fine-tuning, no masked mixture input, and MAR block for recovery loss.
Two variants are considered, left: full fine-tuning on all modules. Right: fine-tuning the adapter only and freezing all other modules.}
 \label{fine-tuning}
\end{figure*}

Notably, we would like to propose a general framework for enhancing existing pre-trained models through incorporating contextual and confidence information. Our approach is designed to be model-agnostic, therefore, we will not discuss the pre-training stage, as it can be any conventionally trained model.

To augment existing pre-trained models with contextual and confidence knowledge, we introduce a two-stage fine-tuning method. In the first stage, which we call global fine-tuning, we enhance the pre-trained vanilla AV-TSE model using the MAR strategy to incorporate contextual information into the separator, improving overall extraction performance.
As for the second stage, also called the ``confidence-aware fine-tuning'' stage, we utilize the pre-trained FCS model to identify unreliable extraction segments and address these hard samples through both self-supervised and supervised strategies. We will detail both stages in the following sections and present the complete procedure in Algorithm \ref{algorithm2}.

\subsection{First Stage: Global Fine-Tuning}
\label{sec:gf}

In the global fine-tuning stage, we aim to enhance overall extraction performance by leveraging the proposed MAR strategy. We term this stage ``global" because the masked segments are randomly selected across the temporal dimension of each utterance, illustrated in Fig. \ref{fine-tuning}(a).


As defined in Equation~\ref{eq:loss_stage1}, the loss function in this stage comprises three components, where $L_{SI-SDR}(\hat{y}, y)$ represents the Scale-Invariant Signal-to-Distortion Ratio (SI-SDR) loss between the extracted output and ground truth.

\begin{align}
 \label{eq:loss_stage1}
 L &= \theta \cdot L_{\text{MSE}_m}(\hat{X}^m, Y^m) 
 + \delta \cdot L_{\text{MSE}_{um}}(\hat{X}^{um}, Y^{um}) \notag \\
 & + \lambda \cdot L_{\text{SI-SDR}}(\hat{y}, y), 
\end{align}

where,
\begin{align}
 &L_{\text{SI-SDR}}(\hat{y}, y) = -10 \log_{10}\left(\frac{\left\|\frac{\langle \hat{y}, y \rangle y}{\|y\|^2}\right\|^2}{\|\hat{y} - \frac{\langle \hat{y}, y \rangle y}{\|y\|^2}\|^2}\right), \\
 &\hat{X}^{m} = \hat{X} \odot I, \quad Y^{m} = Y \odot I, \\
 &\hat{X}^{um} = \hat{X} \odot \overline{I}, \quad Y^{um} = Y \odot \overline{I}.
 \label{sisdr}
\end{align}

 $L_{MSE_m}(\hat{X}^{m},Y^{m}), L_{MSE_{um}}(\hat{X}^{um},Y^ {um})$ denote the Mean Square Error (MSE) of the masked region and unmasked region, respectively. $I$ and $\overline{I}$ denote mask and inverse mask of extracted speech embedding $X$, respectively. 
 
\vspace{-5pt}
\subsection{Second Stage: Confidence-Aware Fine-Tuning}
\label{sec:cf}
The second fine-tuning stage builds upon checkpoints from the first stage and focuses on improving the most challenging extraction regions. During this confidence-aware fine-tuning stage, the model will focus on the unreliable segments predicted by the FCS model. Specifically, the pre-trained FCS model will be frozen and employed as a hard sample estimator to identify unreliable segments in the AV-TSE outputs.

We propose two strategies to refine these identified unreliable segments:
\begin{enumerate}
 \item A self-supervised strategy that masks the corresponding mixture segments in the input and leverages the MAR strategy to reconstruct them using contextual information.
 \item A supervised strategy that increases the loss penalty for the worst extraction segments
\end{enumerate}


While the supervised strategy explicitly enhances the signal loss in problematic regions, the self-supervised approach compels the model to refine these segments using contextual cues learned through masking and recovery.

\subsubsection{Self-Supervised Fine-Tuning Strategy}
In the self-supervised fine-tuning approach, illustrated in Fig. 
 \ref{fine-tuning}(b), we employ a two-pass process for each mixture speech input. In the first pass, we feed the complete mixture speech through the AV-TSE model and use the frozen FCS model to identify regions with poor extraction quality. In the second pass, we mask these identified regions in the input mixture and apply the MAR strategy for fine-tuning, compelling the AV-TSE model to improve its performance on these challenging segments through contextual learning. 
 
The main difference between the second pass and vanilla MAR strategy lies in the selection of masked regions. While the vanilla MAR strategy employs a random masking strategy of the input speech mixture, our second pass specifically masks regions identified as having poor extraction quality. The loss function remains identical to that used in the first fine-tuning stage.

\subsubsection{Supervised Fine-Tuning Strategy}
For the supervised fine-tuning strategy, we apply increased loss penalties to the poorly extracted segments. To avoid the risk of overfitting and explore the most efficient fine-tuning method, we further explored two different settings. 
\begin{enumerate}
 \item Standard full fine-tuning, where all modules except the visual encoder and FCS model remain trainable.
 \item Adapter-based fine-tuning, where we freeze all modules and only train a newly inserted refinement adapter between the MAR block and decoder, designed to address residual errors not resolved in the first fine-tuning stage.
\end{enumerate}

Both configurations are illustrated in Fig. \ref{fine-tuning}(c). The loss function in this setup is calculated as the weighted sum of $L_{\text{SI-SDR}}$ from the identified problematic regions (masked) and the remaining normal regions (unmasked).


\begin{equation}
L = \theta \cdot L_{\text{SI-SDR}_m}(\hat{y}^m,y^m)+\delta \cdot L_{\text{SI-SDR}{um}}(\hat{y}^{um},y^{um})
\end{equation}

\section{Experimental setup}
\subsection{Baselines}
In this paper, we employ six widely used AV-TSE models for experiments. The architectures of these models could be summarized as follows.
\subsubsection{TDSE}
TDSE \cite{tdse} is a time-domain audio-visual speech separation model that extends TasNet \cite{convtas} to audio-visual scenarios. Multiple TasNet-based mask estimators are cascaded as the extraction module. Additionally, the author also explores various audio-visual fusion positions to find the optimal configuration.
\subsubsection{USEV} 

USEV \cite{usev} is specifically designed to handle target and interfering speakers in various overlapping scenarios.
Considering the absence or presence of the target speaker and the interfering speaker in the current frame, it categorizes the speaker overlapping scenarios into four types. To address these scenarios with a unified approach, it proposes scenario-aware differentiated loss. Moreover, USEV employs Dual-path RNN (DPRNN) blocks~\cite{luo2020dual} as the extraction module, which successfully extends DPRNN to audio-visual scenarios.
\subsubsection{MuSE}

MuSE \cite{muse} is an efficient AV-TSE model jointly optimized with a speaker verification task. MuSE adopts iterative TasNet-based mask estimators as the extraction module. It also integrates iterative speaker encoders to estimate one-the-fly speaker embedding. By optimizing the extraction model with a speaker verification loss, MuSE aims to alleviate the target speaker confusion problem \cite{target-speaker-confusion}.
\subsubsection{AV-HuMAR} 
AV-HuMAR \cite{wu2024target_cvpr} also utilizes iterative TasNet-based mask estimators and incorporates a pre-trained AV-HuBERT 
 \cite{AVhubert} into the cue encoder module to obtain robust self-supervised audio-visual synchronization cues. Additionally, it proposes a MAR strategy to inject both inter- and intra-modality context into the extraction module.
\subsubsection{ImagineNET}
ImagineNET \cite{ImagineNET} is tailored for scenarios involving target visual cue occlusion. Specifically, it employs multiple visual decoders to recover the visual embeddings conditioned on the intermediate estimated target speech. By recovering target visual cues, ImagineNET maintains comparable performance across different degrees of visual occlusion, which is more suitable to real-life scenarios.
\subsubsection{AV-Sepformer} 
AV-Sepformer \cite{av-sepformer} extends Sepformer \cite{sepformer} to audio-visual scenarios. It introduces an effective positional encoding mechanism to align fine-grained audio and visual cues without down-sampling, avoiding potential information loss. Moreover, leveraging the long contextual modeling capability of the transformer-based separator, AV-Sepformer demonstrates significant improvements in extraction performance.
\subsection{Dataset}
\subsubsection{Dataset for AV-TSE Models}
 To train and evaluate the performance of AV-TSE models, we simulate a two-speaker mixture dataset from the VoxCeleb2 \cite{voxceleb2}. Similar to previous work \cite{muse, usev, av-sepformer, ImagineNET,wu2024target_cvpr}, $48,000$ utterances from $800$ speakers are selected for the training set and $36,237$ utterances from 118 speakers are selected for the test set. More specifically, we simulate $20000$, $5,000$, and $3,000$ utterances for the training set, validation set, and test set, respectively. Each target speech utterance has been mixed with an interfering speech utterance at a random Signal-to-Noise ratio (SNR) between $-10$ dB and $10$ dB. The speech sampling rate is $16,000$Hz and the video frame is $25$ FPS. Note that compared to the training set and validation set, all speakers in the test set are unseen speakers. Additionally, all the utterances are clipped to $4$ seconds during training and $4-6$ seconds during inference. 
 
 During the first fine-tuning stage, a random segment of each utterance in the training set is masked with zero values. In the second stage, the location of the masked segment is based on scores predicted by the FCS model. Previous research \cite{wu2024target_cvpr} identified that a $300$ ms mask segment optimizes AV-TSE performance. Accordingly, we set the mask segment duration to $300$ ms in this study.
 

\begin{algorithm}
\caption{Fine-Tuning Strategy}
\label{algorithm2}
\begin{algorithmic}[1]
\State \textbf{Input:} $x\in \mathbb{R}^{1\times T},y\in \mathbb{R}^{1\times T},v\in \mathbb{R}^{1\times T''},G\in (0,T)$
\vspace{4pt}
\hrule height 0.5pt
\vspace{3pt}
\Function{Find\_Worst\_Segment\_Index}{$\hat{y},G$}
 \State $score \gets \text{FCS}(\hat{y})$
 \State $w \gets G \cdot (len(score)/T)$ 
 \State $t \gets (G/w) \cdot \arg\min_i \left( \frac{1}{w} \sum_{j=i}^{i+w-1} \text{score}[j] \right)$
 for $i \in (0,len(score) - w + 1)$, 
 \State \textbf{return} $t, t+G$

\EndFunction
\vspace{4pt}
\hrule height 0.5pt
\vspace{3pt}
\text{Global Fine-Tuning:}
\vspace{3pt}
\hrule height 0.5pt
\vspace{4pt}
\State $\text{AVTSE\_MAR.load\_state\_dict(AVTSE})$
 \State $t \gets \text{random.uniform}(0, T-G)$
 \State $x[:,t:t+G] \gets 0$
 \State $\hat{y}, \hat{X}, Y, I, \overline{I} \gets$ AVTSE\_MAR($x,v,y$)

  \State $\hat{X}^{m} \gets I \odot \hat{X}, Y^{m} \gets I \odot Y$
 \State $ \hat{X}^{um} \gets\overline{I} \odot \hat{X}, Y^{um} \gets \overline{I} \odot Y$
 
 \State $L$ $\gets$ 
 $\theta \cdot L_{MSE_m}(\hat{X}^m,Y^m) + \delta \cdot L_{MSE_{um}}( \hat{X}^{um},Y^{um} ) + \lambda \cdot L_{SI-SDR}(\hat{y},y)$
 
\vspace{4pt}
\hrule height 0.5pt
\vspace{3pt}
\text{Confidence-Aware Fine-Tuning:}
\vspace{3pt}
\hrule height 0.5pt
\vspace{4pt}
\State \text{FCS.load\_state\_dict(FCS)}
\State \text{FCS parameters are frozen}

\If{\textbf{self-supervised fine-tuning strategy}}
\State \text{AVTSE\_MAR.load\_state\_dict(AVTSE\_MAR)}
 \State \textbf{with torch.no\_grad():}
 \State \hspace{0.5cm} $\hat{y} \gets \text{AVTSE\_MAR}(x,v)$
 
 \State $t, t+G$ $\gets$ Find\_Worst\_Segment\_Index($\hat{y}$) 
 \State $x[:,t:t+G] \gets 0$
 \State $\hat{y}, \hat{X},Y,I,\overline{I}$ $\gets$ AVTSE\_MAR($x,v,y$)
 \State $\hat{X}^{m} \gets I \odot \hat{X}, Y^{m} \gets I \odot Y$
 
 \State $ \hat{X}^{um} \gets \overline{I} \odot \hat{X}, Y^{um} \gets \overline{I} \odot Y$
 
 \State $L$ $\gets$ $\theta \cdot L_{MSE_m}(\hat{X}^m,Y^m) + \delta \cdot L_{MSE_{um}}( \hat{X}^{um},Y^{um} ) + \lambda \cdot L_{SI-SDR}(\hat{y},y)$
\EndIf

\If{\textbf{supervised fine-tuning strategy}}
\If{\textbf{only fine-tune adapter}}
\State \text{AVTSE\_MAR\_ADA.load\_state\_dict(AVTSE\_MAR)}
\State AVTSE\_MAR parameters are frozen
\State $\hat{y}$ $\gets$ AVTSE\_MAR\_ADA($x,v,y$)
 \State $t,t+G$ $\gets$ Find\_Worst\_Segment\_Index($\hat{y}$) 
 \State $y^m \gets y[:,t:t+G],\hat{y}^m \gets \hat{y}[:,t:t+G]$
 \State $y^{um} \gets y[:,0:t;t+G:], \hat{y}^{um} \gets \hat{y}[:,0:t;t+G:]$
 
 \State $L\gets
 \theta \cdot L_{SI-SDR_m}(\hat{y}^m,y^m)+\delta \cdot L_{SI-SDR_{um}}(\hat{y}^{um},y^{um})$
\EndIf

\If{\textbf{full fine-tune}}
\State \text{AVTSE\_MAR.load\_state\_dict(AVTSE\_MAR)}
 \State $\hat{y}, \hat{X}, Y,\_,\_$ $\gets$ AVTSE\_MAR($x,v,y$)
 \State $t,t+G$ $\gets$ Find\_Worst\_Segment\_Index($\hat{y}$)
 
 \State $y^m \gets y[:,t:t+G],\hat{y}^m \gets \hat{y}[:,t:t+G]$
 \State $y^{um} \gets y[:,0:t;t+G:], \hat{y}^{um} \gets \hat{y}[:,0:t;t+G:]$
 
 \State $L \gets \theta \cdot L_{SI\text{-}SDR_m}(\hat{y}^m, y^m) + \delta \cdot L_{SI\text{-}SDR_{um}}(\hat{y}^{um}, y^{um})$
\EndIf
\EndIf
 
\end{algorithmic}
\end{algorithm}

\subsubsection{Dataset for FCS model}
As mentioned in the previous section, we simulated TSE output dataset to train the FCS model. Since most previous AV-TSE models test on VoxCeleb2 \cite{muse, usev, av-sepformer, ImagineNET,wu2024target_cvpr}, we also use VoxCeleb2, to simulate the TSE output dataset. However, the VoxCeleb2 corpus is collected from YouTube, which contains much background noise or music. Such interfering noise is not labeled in the original dataset, which may be confused with the unreliable segments we intend to simulate. As aforementioned, we only focus on the unreliable extraction segments with interfering speakers in this paper. To filter out the undesired interfering noise, we follow the dataset pre-processing methods leveraged in AV2Wav, which uses a neural quality estimator to filter out the noisy audios \cite{av2wav}. 
 Specifically, we use \cite{torchaudio_squim_package} to filter out the utterances with predicted SI-SDR (P-SI-SDR) below 23 and predicted PESQ (P-PESQ) below 2.0. From this filtered set, we simulated 20,000 TSE output utterances using our proposed simulation algorithm. 
 

 
\subsection{Metrics}
 \subsubsection{AV-TSE Models}
 For the evaluation
metrics, we select the SI-SDR \cite{SDR}, the SI-SDR improvement (SI-SDRi), and the signal-to-noise ratio (SDR) \cite{SDR} 
to evaluate the speech quality. We also use the perceptual evaluation of
speech quality (PESQ) \cite{pesq} and the short-term objective 
intelligibility (STOI) \cite{stoi} to evaluate perception quality and comprehensibility. For all metrics, higher values indicate better performance.
\subsubsection{FCS Model}
To better assess the FCS model's performance, we employ a sliding window approach to analyze the predicted confidence scores. Using a window size of $L$ and a stride of 1, we perform a pooling operation across the scores. The segment with the lowest average confidence score within its window is identified as the most unreliable segment.

Note that we did not establish a specific confidence threshold to identify the most unreliable segment, mainly for two key reasons:
\begin{itemize}
 \item {First, our FCS model is designed for real-world TSE scenarios, where prior knowledge about the mixture data distribution and the performance of the black-box AV-TSE system is often unavailable. Given this uncertainty, establishing a fixed confidence threshold is not appropriate.}

 \item {Second, with the proposed confidence-aware fine-tuning strategy, we anticipate a reduction in the proportion of poor extraction segments and an overall improvement in extraction performance. Consequently, a fixed confidence threshold may not be suitable, as the extraction performance of the AV-TSE system could vary during the fine-tuning process.}
\end{itemize}
 
To validate the FCS model's prediction accuracy, we compared three types of segments:
\begin{enumerate}
 \item ``Unreliable": The segment identified by the FCS model as most unreliable.
 \item ``Reliable": A randomly selected segment from regions outside the identified unreliable segment.
 \item ``Random": A randomly selected segment from the entire utterance.
\end{enumerate}

We then calculated the chunk-level SI-SDR for each of the three segment types to verify whether there are statistically significant differences in their performance.

\subsection{Implementation Details}
\subsubsection{AV-TSE Models}
For TDSE, MuSE, AV-HuMAR, and ImagineNET, the model parameters $(N, L, B, H, P, X, R)$ are set to $(256, 40, 256, 512, 3, 7, 4)$. 
For USEV, $(N, L, B, H, K, R)$ are set to $(256,40,64,128,100,6)$. For AV-Sepformer , $(L,C,N_{intra},N_{inter}, N, N_{head})$ are set to $(16,160,8,7,256,8)$. The scale of loss functions $(\theta,\delta,\lambda)$ in the first and second stages are set to $(1,5,1)$. The masked gap $G$ of the mixture speech is set to $300$ ms. $T$ and $T''$ denote waveform duration and corresponding visual cue duration. The architecture of the newly inserted refinement adapter is a single Conv1d layer.


For the first fine-tuning stage, the fine-tuning epoch is set to $50$. For the second fine-tuning stage, the fine-tuning epoch is set to $30$. For each fine-tuning stage, the learning rate is set to $15e-5$, and the checkpoints are selected based on the best validation performance.

\subsubsection{FCS Model}
 We set $\left\{\alpha, \beta,N_{max},g\right\}$ with $\left\{0.9,0.2,20,10\right\}$, where $g$ is in million seconds. Note that we found that changing $g$ and $N_{max}$ around the selected values has no significant impact on the FCS model. We also conducted preliminary experiments to determine the optimal $\left\{\alpha, \beta\right\}$ configuration. 
Specifically, we conducted a grid search for the values of $\alpha$ and $\beta$ ranging from 0 to 1, with a stride of 0.1, across the entire TSE training and validation set. We simulated unclean extraction results with selected $\alpha$ and $\beta$, then calculated the SI-SDR between the simulated target speech extraction and ground truth. Note that the ideal extraction results should approximate point $(1,0)$, indicating the extracted speech is equal to the ground truth. However, most AV-TSE systems could achieve SI-SDR results concentrated in the lower right corner. Thus we want to find a worse extraction state that is worse than all backbones but still around the corner.

\begin{figure}[!htb]
\centering
\includegraphics[scale=0.55]{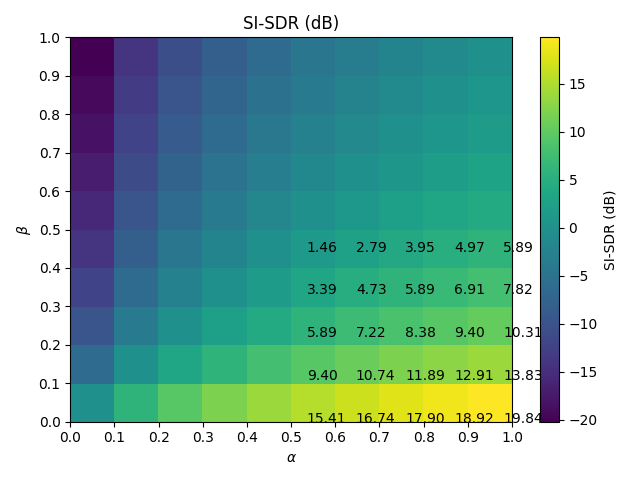}

\caption{Visualization of SI-SDR distribution with grid-search of $\alpha$ and $\beta$}
\label{sisdr_matrix}
\end{figure}

As shown in Fig.\ref{sisdr_matrix}, we observed that setting $\alpha$ to 0.9 and $\beta$ to 0.2 resulted in a simulated speech utterance achieving an SI-SDR of 9.40. This value is lower than the extraction results from the weakest AV-TSE baseline used in this study, which achieved an SI-SDR of 10.725. We selected this configuration because the simulated speech segment represents an extraction state where the target speech is relatively clean but still contains some interfering speech. This indicates that the extraction quality has room for improvement. Based on our modeling of extraction states, we argue that sampling such a point offers generalizability to most AV-TSE systems. Furthermore, we believe that extending this approach to a finer grid search would remain feasible and likely yield more precise configurations.

 In the score prediction model, the speech encoder utilizes a Conv1d layer with a channel size of $256$, a kernel size of $320$, and a stride size of $160$. $T'$ is the duration of the confidence score, obtained after passing the target speech into the Conv1d layer. The FCS predictor comprises $3$ transformer layers with a hidden dimension of $256$ and $4$ heads and $2$ linear layers with channels of $(256,256)$ and $(256,1)$, respectively. We train the model with a learning rate of $1e-5$ for $30$ epochs with a batch size of $40$ and select the checkpoint with the best performance on the validation set.

\begin{table*}[!ht]
\begin{center}
\caption{Comparison of supervised fine-tuning strategy. Here we employ six AV-TSE models as baselines. `FT' denotes fine-tuning parameters. Specifically, `F' denotes fine-tuning all AV-TSE modules excluding Visual Encoder. `A' denotes only fine-tuning the Refine Adapter, all other parameters will be frozen. Note that the pre-train FCS model will always be frozen. 
 'Type' denotes fine-tuning types in confidence-aware fine-tuning, including SS (Self-supervised) and S (Supervised). For SS, the masked gap of the mixture is set to 300 ms.}
\label{supervised-ft-res}
\begin{tabular}{p{0.11\textwidth}p{0.05\textwidth}p{0.07\textwidth}p{0.05\textwidth}p{0.05\textwidth}p{0.08\textwidth}p{0.08\textwidth}p{0.08\textwidth}p{0.07\textwidth}p{0.07\textwidth}} 
\toprule
\textbf{Model} & \textbf{Global} & \textbf{Confidence } & \textbf{FT} & \textbf{Type} & \textbf{SI-SDR $(\uparrow)$} & \textbf{SI-SDRi $(\uparrow)$} & \textbf{SDR $(\uparrow)$} & \textbf{PESQ $(\uparrow)$} & \textbf{STOI $(\uparrow)$} \\
\toprule
\multirow{5}{*}{TDSE} & \ding{55} & \ding{55} & - & -& 10.725 & 10.771 & 11.099 & 2.592 & 0.859 \\
   & \ding{51} & \ding{55} & F & - &11.893 & 11.940 & 12.422 & 2.929 & 0.881 \\
   & \ding{51} & \ding{51} & F & SS& 11.989 & 12.037 & 12.516 & 2.949 & 0.882 \\
   & \ding{51} & \ding{51} & F & S & 12.096 & 12.143 & 12.537 & 2.943 & 0.883 \\
   & \ding{51} & \ding{51} & A & S& 11.944 & 11.991 & 12.420 & 2.929 & 0.881 \\

\toprule
\multirow{5}{*}{USEV} & \ding{55} & \ding{55} & - & - & 10.785 & 10.829 & 11.332 & 2.646 & 0.862 \\
   & \ding{51} & \ding{55} & F & - & 11.387 & 11.435 & 11.877 & 2.776 & 0.871 \\
   & \ding{51} & \ding{51} & F & SS & 11.823 & 11.871 & 12.275 & 2.857 & 0.872 \\
   & \ding{51} & \ding{51} & F & S& 11.706 & 11.753 & 12.142 & 2.801 & 0.877 \\
   & \ding{51} & \ding{51} & A & S& 11.444 & 11.491 & 11.917 & 2.778 & 0.872 \\
   
\toprule
\multirow{5}{*}{MuSE} & \ding{55} & \ding{55} & - & -& 11.458 & 11.506 & 11.836 & 2.706 & 0.873 \\
   & \ding{51} & \ding{55} & F & - & 11.816 & 11.832 & 12.233 & 2.820 & 0.878 \\
   & \ding{51} & \ding{51} & F & SS & 11.907 & 11.954 & 12.342 & 2.862 & 0.878 \\
   & \ding{51} & \ding{51} & F & S & 11.734 & 11.781 & 12.169 & 2.784 & 0.874 \\
   & \ding{51} & \ding{51} & A & S & 11.822 & 11.869 & 12.260 & 2.831 & 0.877 \\
   
\toprule
\multirow{5}{*}{AVHuBERT-TSE} & \ding{55} & \ding{55} & - & -& 11.728 & 11.771 & 12.043 & 2.765 & 0.878 \\
    & \ding{51} & \ding{55} & F & -& 12.331 & 12.379 & 12.726 & 2.922 & 0.887 \\
    & \ding{51} & \ding{51} & F & SS& 12.639 & 12.686 & 13.073 & 3.005 & 0.893 \\
    & \ding{51} & \ding{51} & F & S& 12.538 & 12.586 & 12.953 & 2.975 & 0.891 \\
    & \ding{51} & \ding{51} & A & S& \textcolor{gray}{11.194} & \textcolor{gray}{11.242} & \textcolor{gray}{11.668} & \textcolor{gray}{2.803} & \textcolor{gray}{0.874} \\
    
\toprule
\multirow{5}{*}{AV-Sepformer} & \ding{55} & \ding{55} & - & - & 12.472 & 12.519 & 12.824 & 2.844 & 0.884 \\
    & \ding{51} & \ding{55} & F & - & 12.941 & 12.957 & 13.113 & 3.022 & 0.893 \\
    & \ding{51} & \ding{51} & F & SS& \textbf{13.258} & \textbf{13.305} & \textbf{13.683} & \textbf{3.090} & \textbf{0.898} \\
    & \ding{51} & \ding{51} & F & S& \textbf{13.361} & \textbf{13.408} & \textbf{13.771} & \textbf{3.012} & \textbf{0.896} \\
    & \ding{51} & \ding{51} & A & S & 13.043 & 13.059 & 13.526 & 3.007 & 0.894 \\
    
\toprule
\multirow{5}{*}{ImagineNET} & \ding{55} & \ding{55} & - & -& 12.831 & 12.878 & 13.166 & 2.944 & 0.892 \\
    & \ding{51} & \ding{55} & F & -& 12.961 & 12.977 & 13.359 & 3.026 & 0.898 \\
    & \ding{51} & \ding{51} & F & SS & 13.088 & 13.135 & 13.527 & 3.073 & 0.901 \\
    & \ding{51} & \ding{51} & F & S& 13.135& 13.183 & 13.536 &3.019& 0.899 \\
    & \ding{51} & \ding{51} & A & S & 13.035 & 13.083 & 13.440 & 3.047 & 0.899 \\
    
\bottomrule
\end{tabular}
\end{center}
\end{table*}

\section{Experimental Results}
\subsubsection{Baselines}
Six popular AV-TSE models are trained as our baselines, termed as vanilla pre-trained AV-TSE models
(Vanilla). As shown in Table \ref{supervised-ft-res}, ImagineNET achieves the highest SI-SDR of $12.831$, substantially outperforming other baselines. This superior performance can be attributed to its unique approach of using multiple visual decoders to reconstruct target visual embeddings from both intermediate estimated speech and refined visual cues, likely enabling better cross-modality context capture. AV-Sepformer achieves the second-best performance with an SI-SDR of $12.472$, benefiting from its transformer-based separator's long-range contextual modeling and sophisticated audio-visual fusion mechanism. 

\subsubsection{Performance With Global Fine-Tuning}
We evaluate the vanilla pre-trained AV-TSE model combined with the first stage of global fine-tuning (Vanilla + GF), as described in Section~\ref{sec:gf}. As shown in Tables~\ref{supervised-ft-res}, the proposed Vanilla + GF stage achieves consistent improvements across all six systems, demonstrating the effectiveness of integrating both inter- and intra-modality contextual correlations. Notably, for TDSE systems, an improvement of over 1 dB was achieved in terms of SI-SDR and SI-SDRi.

\subsubsection{Performance With Confidence-Aware Fine-Tuning} To evaluate the second-stage confidence-aware fine-tuning, we present results for the two setups described in Section~\ref{sec:cf}.

\textbf{Self-Supervised Fine-Tuning Strategy} Table \ref{supervised-ft-res} presents the experimental results with the self-supervised fine-tuning strategy. For each AV-TSE model, we compare three configurations: Vanilla, Vanilla + GF, and the vanilla model with both global and confidence-aware fine-tuning (Vanilla + GF + CF).

The results show that despite GF already achieving significant improvements over the baseline, adding CF yields further gains. For example, GF improves AV-Sepformer's SI-SDR from 12.472 to 12.941, while GF + CF further increases it to 13.258. These results confirm our hypothesis that two-stage fine-tuning enables progressive refinement from coarse to fine extraction. Notably, AV-Sepformer + GF + CF achieves the best performance among all AV-TSE systems tested.

\begin{figure*}[t]
\centering
\includegraphics[scale=0.1]{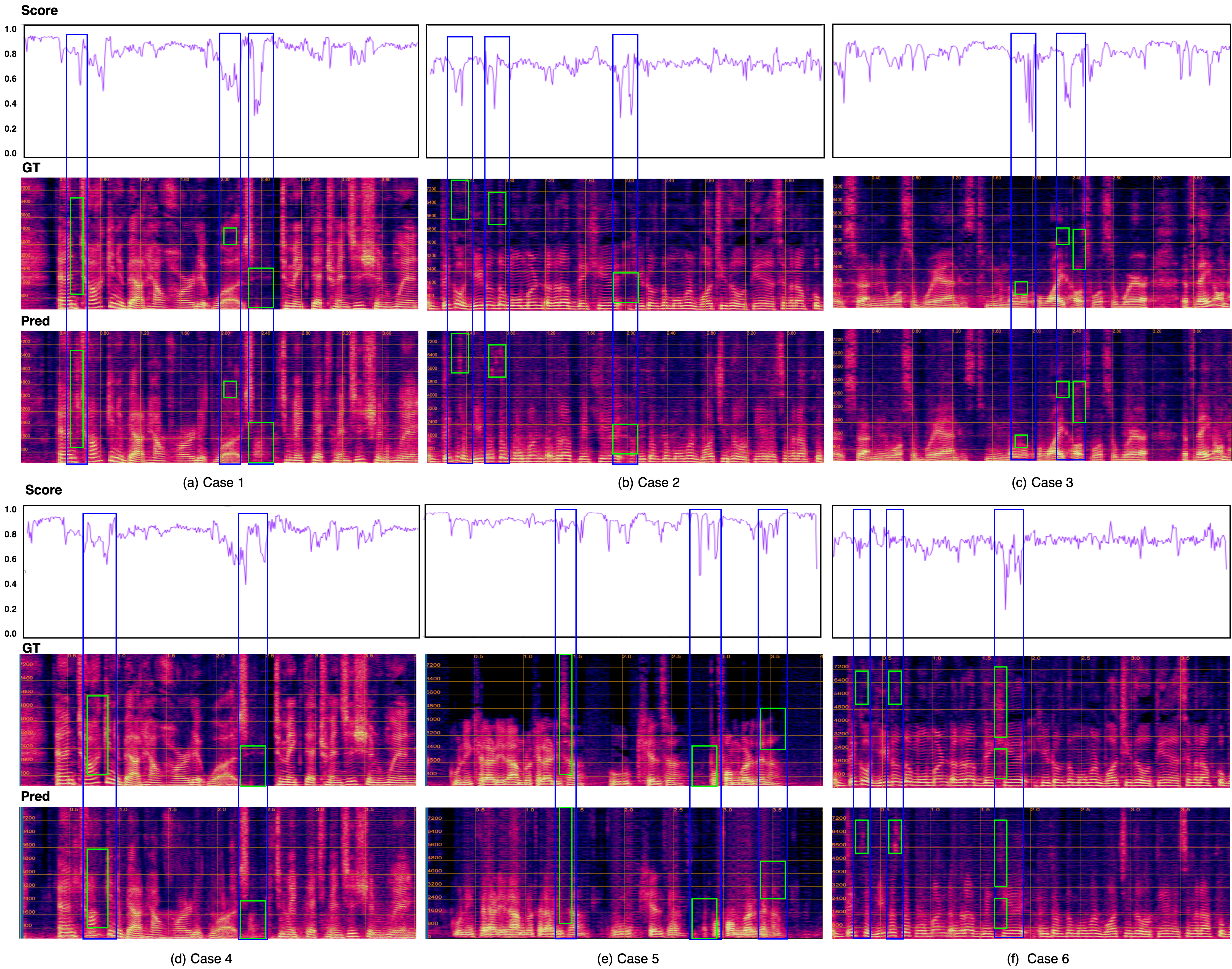}
 \caption{The case study for visualization of captured unreliable segments. For each case, the first line is the corresponding confidence scores predicted from the pre-trained FCS model, the second line is the spectrogram of ground truth, and the third line is the spectrogram of extracted target speech. Each score represents 10 ms waveform and the total duration is 4000 ms.}
\label{score_case_study}
\vspace{-10pt}
\end{figure*}




\textbf{Supervised Fine-Tuning Strategy} As shown in Table \ref{supervised-ft-res}, similar to the self-supervised fine-tuning results, we report the supervised experimental results involving the Vanilla, Vanilla + GF, Vanilla + GF + CF. 
Note that the CF is supervised and without MAR strategy.

Among all the experimental results we observe that AV-Sepformer + GF + CF, utilizing full parameters, achieves the best performance, with an SI-SDR of $13.361$. This is slightly higher than using the self-supervised fine-tuning strategy, which attains an SI-SDR of $13.258$.

Additionally, in the CF setup, full fine-tuning generally outperforms the Refine Adapter-only approach, which is expected given the larger number of trainable parameters. While adapter fine-tuning achieves reasonable improvements over the first stage, it occasionally leads to performance degradation, yielding results inferior to vanilla AV-TSE models like AVHuBERT-TSE. These findings suggest that a single adapter layer may be insufficient for refining challenging samples. In contrast, full fine-tuning provides better refinement for these difficult cases, though at the cost of increased computational resources.

Furthermore, the improvements resulting from the proposed fine-tuning strategies are more pronounced for weaker baselines. For example, in the TDSE system, our best configuration enhances SI-SDR from 10.275 to 12.096, representing an absolute improvement of 1.821 dB. Similarly, USEV shows an absolute increase of 0.921 dB (from 10.785 to 11.706). In contrast, stronger baselines like ImagineNET exhibit relatively modest gains, with SI-SDR improving by only 0.304 dB (from 12.831 to 13.135). Similar patterns emerge in the self-supervised fine-tuning results. Upon analyzing the architectural designs, we found that stronger baselines such as AV-Sepformer and ImagineNET already, to some extent, incorporate sophisticated audio-visual fusion mechanisms that capture high-level contextual information during extraction. Nevertheless, our approach still provides complementary benefits to these robust systems.

\textbf{Comparison of Self-supervised and Supervised Fine-Tuning Strategy}

Here we only compare the self-supervised fine-tuning (SS) strategy and the supervised full fine-tuning (S) strategy, because supervised fine-tuning using an adapter may lead to overfitting due to too few trainable parameters.

 \begin{itemize}
 \item {For SS is better: SS with its long contextual modeling ability could compensate for the local contextual modeling capabilities of DPRNN in USEV. SS effectively extracts useful features from the model, while S may cause MuSE to focus on background noise. This issue arises because MuSE includes a speaker verification module, which can lead to unreliable segments containing more noise and fewer interfering speakers. Additionally, AVHuBERT-TSE employs modality dropout during training, aligning with the masking mechanism, which makes SS more suitable. }

 \item {For S is better: We observe that TDSE, the weakest backbone, performs better with S. Since TDSE does not incorporate additional speaker modeling or contextual modeling modules, S better may be because its training objective is simpler and aligns with the pre-training stage.
 Both AV-Sepformer and ImagineNET demonstrate strong contextual modeling abilities; the former benefits from a transformer-based extractor, while the latter employs a hierarchical interlace extractor. Thus, context information has been thoroughly utilized during the pretraining or GF stage and shows only limited effectiveness during the CF stage. As a result, relying on stronger supervised signals may yield better outcomes.}
\end{itemize}

\subsubsection{FCS Model Performance}
To verify the prediction accuracy of the pre-trained FCS model, we analyzed the chunk-level SI-SDR of segments identified as unreliable by the model, as shown in Table \ref{real}. We tested the model using different chunk sizes ($L$) of 200 ms, 400 ms, and 600 ms to assess its robustness across various temporal resolutions.


\begin{table}[!htb]
\caption{\centering Comparison of average SI-SDR from the random, reliable, and unreliable regions of \textbf{unseen AV-TSE test results}.}
\label{real}
\begin{tabular}{p{0.13\textwidth}p{0.06\textwidth}p{0.05\textwidth}p{0.05\textwidth}p{0.05\textwidth}}
\toprule
Model & Segment & 200(ms) & 400(ms) & 600(ms) \\
\toprule
\multirow{3}{*}{TDSE } & Random & 12.366& 10.862 & 11.523 \\
 & Reliable & 12.375 & 11.055 & 12.065\\
 & Unreliable &\textbf{9.769} & \textbf{8.148} & \textbf{9.068} \\
 \hline
\multirow{3}{*}{USEV } & Random &10.182 & 10.802 & 10.891\\
 & Reliable & 10.659 & 11.083 & 11.237 \\
 & Unreliable & \textbf{8.149} & \textbf{8.458} & \textbf{8.711} \\
 \hline
\multirow{3}{*}{MuSE } & Random & 10.864 & 11.377 &11.495 \\
 & Reliable & 11.235 &11.901 &12.050 \\
 & Unreliable & \textbf{8.539}& \textbf{8.828} & \textbf{9.0730}\\
 \hline
 \multirow{3}{*}{ImagineNET } & Random & 12.158 & 12.501 & 12.627\\
 & Reliable & 12.403 & 12.879 & 13.104 \\
 & Unreliable & \textbf{9.769} & \textbf{10.051} & \textbf{10.336}\\
 \hline
 \multirow{3}{*}{AV-Sepformer } & Random & 12.109 & 12.570 &12.813 \\
 & Reliable & 12.214 & 12.740 & 13.088 \\
 & Unreliable & \textbf{9.730}& \textbf{10.054} & \textbf{10.390} \\
 \hline
\toprule
\end{tabular}
\vspace{-10pt}
\end{table}

\begin{figure}[!htb]
\centering
\includegraphics[scale=0.55]{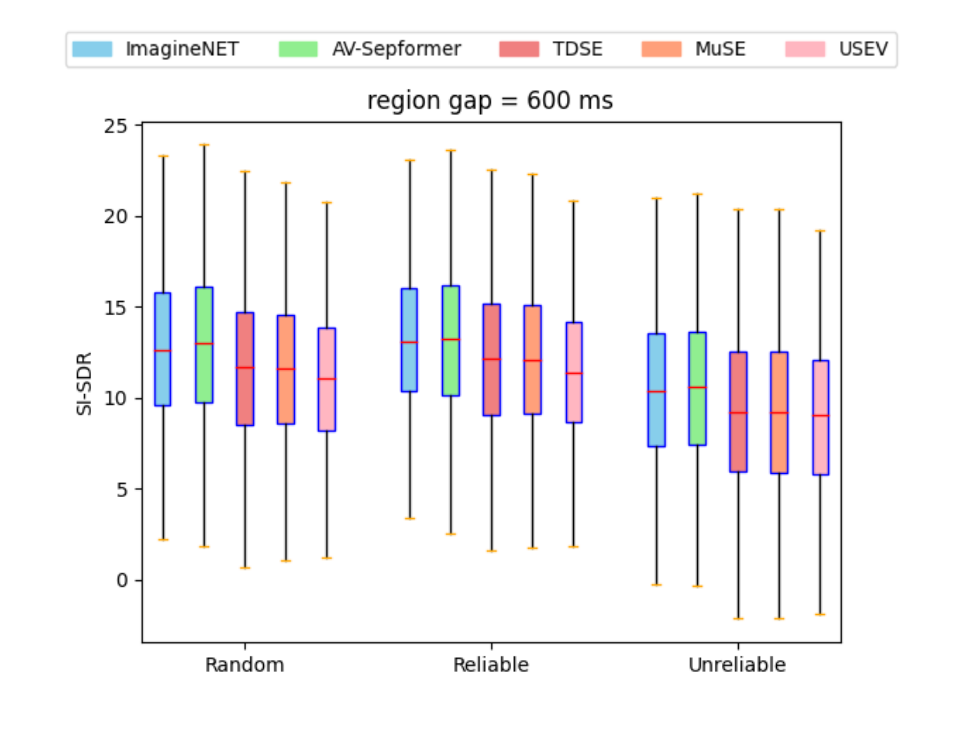}
\vspace{-20pt}
\caption{Visualization of SI-SDR distribution from the reliable, unreliable, and full regions of \textbf{unseen AV-TSE test results}. Here we report the region gap with $600$ ms. Here we employ five baselines denoted in different colors.}
\label{box}
\end{figure}

Table \ref{real} demonstrates that segments identified as unreliable consistently show significantly lower SI-SDR values compared to both reliable and randomly selected segments, while the reliable segments achieve the highest SI-SDR values. These observations validate the effectiveness and accuracy of the proposed FCS model. Additionally, we observe that as the segment duration increases, the SI-SDR gap between reliable and unreliable segments also widens. Such a trend occurs because the well-extracted segment increases, so the SI-SDR of the reliable segment also increases. In versa, the unreliable segment becomes worse and SI-SDR is lower. To make a clear comparison, we visualize $600$ ms segment results, as shown in Fig. \ref{box}.

\begin{figure*}[!htb]
\centering
\includegraphics[scale=0.12]{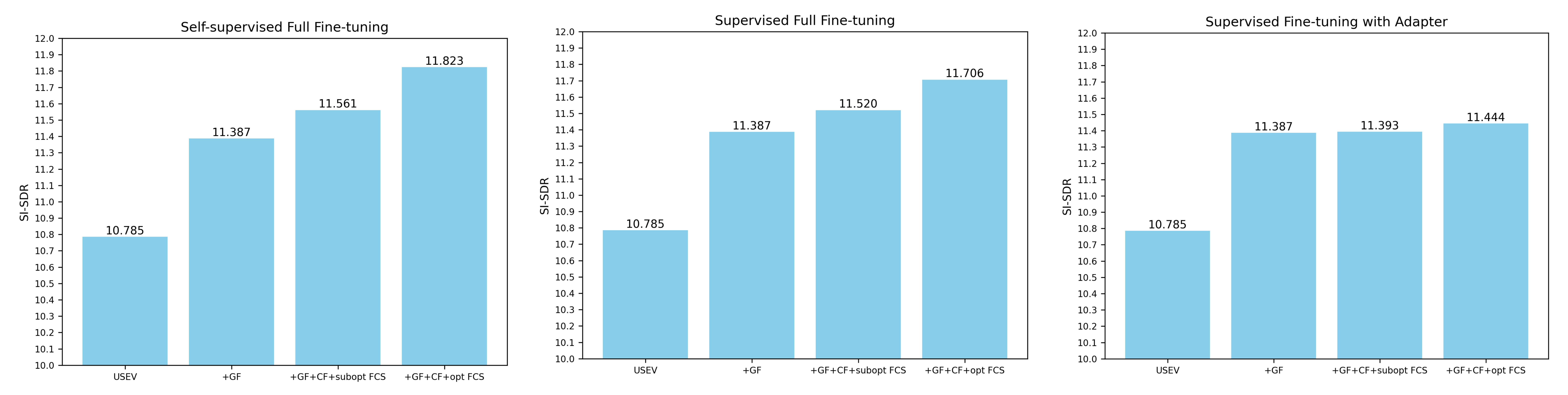}
\caption{SI-SDR comparison of USEV+GF+CF using a suboptimal FCS model and using the optimal FCS model. For each figure, from left to right: vanilla USEV, vanilla USEV+GF, vanilla USEV+GF+CF with suboptimal (subopt) FCS model, vanilla USEV+GF+CF with optimal (opt) FCS model.}
\label{subopt_fcs}
\end{figure*}
\subsubsection{Case Study For FCS Model}
To demonstrate the effectiveness of our approach, we visualize the spectrograms of identified unreliable segments from the AV-TSE results. As presented in Fig. \ref{score_case_study},
for each case, we present three rows: the predicted fine-grained confidence scores of extracted target speech (top), the ground truth target speech (middle), and the extracted target speech (bottom). Blue rectangles align the low confidence score region with the corresponding spectrograms, while green rectangles highlight specific extraction errors compared to the ground truth.
 
Across all six cases, noticeable discrepancies between predictions and ground truths can be observed both in low-frequency and high-frequency parts, all corresponding to areas with low confidence scores. It is also worth noting that cases $1$ and $4$ are derived from the same mixture. However, case $1$ is extracted from TDSE, while case $4$ is extracted from TDSE + GF. The SI-SDR in case $1$ achieves $10.122$, whereas in case $4$ achieves $11.833$, showing an improvement of $1.71$. Correspondingly, the confidence score of case $4$ is higher than case $1$, indicated by the blue rectangle in Fig. \ref{score_case_study}. Additionally, the amount of false extractions highlighted in the green rectangle is lower in case $4$ compared to case $1$. These findings highlight the effectiveness of the FCS model in hard sample mining.

\subsubsection{How Does FCS Model Performance Affect AV-TSE Performance?}
To verify whether FCS model performance affects AV-TSE performance, we used a suboptimal FCS model in the CF stage. Specifically, we selected the FCS model pre-trained for only one epoch, representing an underfitting FCS model with insufficient prediction ability. For AV-TSE backbones, we selected USEV, which demonstrated the most significant SI-SDR relative improvement on average of all fine-tuning configurations in the CF stage.

The test set performance after the CF stage is presented in Fig \ref{subopt_fcs}. Compared to fine-tuning with a well-trained FCS model, using an underfitting FCS model for fine-tuning led to a reduction in SI-SDR across all three strategies involving FCS-based confidence prediction.

This finding suggests that the underfitting FCS model may lead to poorer prediction for unreliable extraction segments, thereby, the training objective in the CF stage may not be optimal. Consequently, the final extraction performance is not as good as using a well-trained FCS model. However,  even with a suboptimal training objective, we observe that it may not lead to a worse extraction performance than vanilla AV-TSE. Instead, it may produce a suboptimal extraction result with non-significant improvements.



\subsubsection{Validation of Visual Impaired Scenario}

To validate the robustness of the proposed two-stage fine-tuning approach when target visual cues are impaired, we simulate two most common scenarios:
\begin{itemize}
 \item Visual Occlusion: Target visual cues occluded by certain obstacles.
 \item {Low Resolution: Poor lighting conditions or low camera resolution.}
\end{itemize}

\begin{figure}[htbp]
 \centering
\includegraphics[scale=0.7]{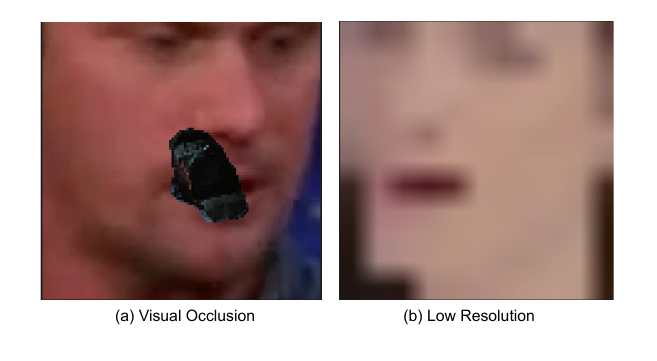}
\vspace{-20pt}
 \caption{\centering The two common visual impairment conditions in real-world scenarios include visual occlusion and low resolution.}
 \label{vocc_type}
\end{figure}

In Fig. \ref{vocc_type}, we present two simulated data samples with target visual cue impaired.

\begin{table}[!ht]
\begin{center}
\caption{Evaluation with 2 types of visual impaired conditions on Voxceleb2 test set, GF and CF denote global and confidence. Occ, Low, and Clean denote visual occlusion, low resolution, and visual clean scenarios, respectively.}
\vspace{-10pt}
\label{VOCC}
\begin{tabular}{p{0.075\textwidth}p{0.01\textwidth}p{0.01\textwidth}p{0.01\textwidth}p{0.02\textwidth}p{0.05\textwidth}p{0.06\textwidth}p{0.07\textwidth}} 
\toprule
\textbf{Model} & \textbf{\centering GF} & \textbf{\centering CF} & \textbf{FT} & \textbf{Type} &\textbf{Occ $(\uparrow)$} & \textbf{Low $(\uparrow)$} &\textbf{Clean $(\uparrow)$}\\
\toprule

\multirow{5}{*}{TDSE} & \ding{55} & \ding{55} & - & - &9.17 & 8.55 & 10.73 \\
   & \ding{51} & \ding{55} & F & - & 9.69 & 8.87 &11.89 \\
   & \ding{51} & \ding{51} & F & SS &\textbf{9.89} &\textbf{9.06} &11.99 \\
   & \ding{51} & \ding{51} & F & S &9.87 & 8.86 & \textbf{12.10} \\
   & \ding{51} & \ding{51} & A & S &9.73 & 8.79 & 11.94 \\  
\toprule
 
\multirow{5}{*}{AV-Sepformer} & \ding{55} & \ding{55} & - & - &11.26 & 10.72 & 12.47 \\
   & \ding{51} & \ding{55} & F & - & 11.64 & 11.06 & 12.94 \\
   & \ding{51} & \ding{51} & F & SS &12.03 &11.58 & 13.26 \\
   & \ding{51} & \ding{51} & F & S &\textbf{12.14} &\textbf{11.58} & \textbf{13.36} \\
   & \ding{51} & \ding{51} & A & S &11.74 &11.17 & 13.04 \\

\bottomrule
\end{tabular}
\end{center}
\end{table}

We select the strongest backbone (AV-Sepformer) and the weakest backbone (TDSE) for our analysis. It is important to note that we do not retrain or fine-tune our model on any dataset with impaired visual cues; instead, we directly evaluate our model on the simulated visual cue impaired test set.
Table~\ref{VOCC} reports the SI-SDR results of visual occlusion, low resolution, and visual clean results. For both TDSE and AV-Sepformer, the visual impairment SI-SDR is lower than under visual clean conditions, which is reasonable due to challenging visual cues and domain mismatch. However, with the proposed two-stage fine-tuning strategy, performance consistently improves compared to the vanilla backbones.
Considering none of our models were exposed to visual impairments, the proposed two-stage fine-tuning methods still perform well on the visual impairments test set, demonstrating that contextual and confidence information play important roles in the extraction and refinement of unseen scenarios, which is valuable for real-world applications.

\section{Conclusion}
In this study, we introduce $C^2$AV-TSE, which incorporates context and confidence information into AV-TSE models. The proposed MAR strategy assists AV-TSE models in learning inter- and intra-modality contextual correlations during extraction, while the FCS model can be leveraged to detect unreliable extraction segments from TSE output. Furthermore, the proposed two-stage fine-tuning strategy successfully integrates the MAR strategy and the FCS model, demonstrating stable performance improvements across six AV-TSE models. In the future, we would like to explore more detailed reasons for performance improvements and apply the fine-tuning strategy to more complex scenarios.


\bibliographystyle{IEEEtran}
\bibliography{refs} 

\end{document}